# Predicting Privacy Attitudes Using Phone Metadata

Isha Ghosh and Vivek K. Singh

School of Communication and Information, Rutgers University, New Brunswick, U.S.A.
{isha.ghosh, v.singh}@rutgers.edu

**Abstract.** With the increasing usage of smartphones, there is a corresponding increase in the phone metadata generated by individuals using these devices. Managing the privacy of personal information on these devices can be a complex task. Recent research has suggested the use of social and behavioral data for automatically recommending privacy settings. This paper is the first effort to connect users' phone use metadata with their privacy attitudes. Based on a 10-week long field study involving phone metadata collection via an app, and a survey on privacy attitudes, we report that an analysis of cell phone metadata may reveal vital clues to a person's privacy attitudes. Specifically, a predictive model based on phone usage metadata significantly outperforms a comparable personality features-based model in predicting individual privacy attitudes. The results motivate a newer direction of automatically inferring a user's privacy attitudes by looking at their phone usage characteristics.

## 1 Introduction

Recent results have pointed to a significant awareness of the dangers of sharing information and pictures on social networking sites (SNSs), and young adults today are careful about constructing identity and online information disclosure [6, 7]. However, the fact that a number of interactions happen over smartphones which constantly receive and send out data signals containing personal information often slips under the radar [17], [21, 22]. Enormous amounts of personal data are being captured from user's smartphones to personalize user requirements however, there is little work done to understand how this mobile metadata (particularly call logs,) can be used to build personalized privacy settings for users. In this paper, we propose pivoting the use of phone metadata (particularly call logs) to allow users to utilize their own data to obtain personalized privacy recommendations.

As a first step toward this goal, we test the ability of an individuals' phone metadata to predict user's privacy attitudes. Once established, such interconnections could be used to automatically define user's privacy settings without the need for manual surveys or weaving through complicated choices.

This paper makes two important contributions.

1. Motivates and grounds the use of phone metadata as a method of assessing user privacy needs.
2. Lays the groundwork for building automated models that define user's privacy attitudes, without the need for explicit surveys.

## 2    Related work

We focus the presentation of related work on research projects that discuss information sharing within and outside communities and their impact on privacy attitudes. As more and more aspects of human life get mediated by mobile phones, a quick and easy way to identify privacy attitudes for users may be useful to suggest default settings and configurations in a variety of applications and scenarios faced by the user. However, in order to effectively suggest such settings and configurations, it is important to gain an understanding of an individual's privacy attitudes.

There have been several attempts to define privacy attitudes. In a systematic discussion of the different notions of privacy Introna and Poloudi (1999) developed a framework of principles that explored the interrelations of interests and values for various stakeholders where privacy concerns have risen. The central idea around an individual's privacy attitude is the desire to keep personal information out of the hands of others, along with the ability to connect with others without interference. In this context, concern for privacy is a subjective measure—one that varies from individual to individual based on that person's own perceptions and values. In other words, different people have different levels of concern about their own privacy [18].

However, a concern for privacy does not translate into similar behavior. Previous research [1, 2], [27] has explored the dichotomy that exists between concerns about privacy and actual behaviors exhibited by individuals. The focus of this study is to understand the attitude and concerns of individuals towards privacy and how these concerns influence (or are influenced by) their real-world social behavior.

While there exist a number of studies to measure privacy attitudes and behaviors in online interactions, [1], [27], [30], we wanted to get a sense of privacy attitudes in both online and offline behaviors. Therefore we use the Westin's [32, 33, 34, 35, 36, 37, 38] studies to gain a holistic understanding of privacy concerns exhibited by individuals'. While there have been multiple recent efforts toward building newer measures of privacy, [1, 2], [7] Westin's work remains one of the most comprehensive approaches towards obtaining a well-rounded understanding of privacy attitudes exhibited by an individual. This paper uses the Privacy Segmentation Index [38] that categorizes individuals' into one of three categories based on their levels of privacy concerns:

- **Fundamentalists:** who feel very strongly about privacy and grant it an especially high value
- **Pragmatists:** who also have strong feelings about privacy but can also see the benefits from surrendering privacy in situations where they believe care is taken to prevent the misuse of this information; and
- **Unconcerned:** those who have no real concerns about privacy or about how other people and organizations are using information about them.

Previous research has explored various methods for improving the understanding of complex privacy settings [14], [23], [31] in SNSs. A recent study automatically generates privacy settings for any images uploaded by the user [25]; another study describes a privacy wizard for SNSs that describes a particular user's privacy preferences based on a limited amount of user input [15]. While this research work can help in preserving privacy in online SNSs, there is a gap in the literature around using cell phone metadata to generate automated privacy settings for individuals. Our research focuses on using this cellphone metadata as a predictor of an individual's privacy attitudes.

## 3     Study Undertaken

We adopt the Privacy Segmentation Index (PSI) to gain an understanding of privacy attitudes displayed by individuals. This survey consisted of statements designed to measure levels of concern about personal information disclosed by individuals to companies or businesses and their concerns about whether their information was being protected or not. Each of these questions required responses on a 4-point scale ranging from strongly disagree to strongly agree and individuals were classified as Fundamentalists, Pragmatists or Unconcerned based on their responses.

Participants for this study were recruited from Rutgers University, New Brunswick. During the study, participants were invited to the study site to read and sign the consent form and fill out an online survey. The survey consisted of the Privacy Segmentation Index [38], and demographic questions (e.g. gender, age). While the participants completed the survey, they were asked to install the study client on their phones. The study client collected call logs, sms logs, and location logs, over a 10 week study period. Cellphone metadata collected by the study client and the participant answers to the privacy attitude and behavior surveys were analyzed to test multiple hypotheses. A total of 53 participants completed the study i.e. installed the study client app and completed the privacy survey. Of these 31 (59%) were men and 18 (34%) were women (demographic data was unavailable for three participants). The majority of participants were undergraduates between the ages of 18-21 years.

Based on the information collected from this app we defined a set of features that will allow us to explore relationships between mobile phone interactions and privacy attitudes. Below are the variables and their definitions:

**Table 1.** Variables used in the study

| Variable Name | Definition |
|---|---|
| Privacy Concern (Output Variable) | Score as determined by responses to Westin's Privacy Segmentation Index. Scores for each question on a Likert Scale of 1-4 were added together to get a combined Privacy Concern Score. |
| Call Count | $n(Calls)$<br>Total number of calls received or made by participants in the duration of the study |
| Call Duration | $\sum(time\ spent\ on\ calls)$<br>Sum of time spent on all calls received or made in the duration of this study |
| Missed Call Rate | *(Number of missed calls / Call Count) * 100*<br>This is percentage of calls missed (not answered) by the participant. |
| Call Response Rate | *(Number of responded missed calls/ Number of missed calls) * 100*<br>Where "responded missed calls" are those which were returned within 1 hour of the missed call. |
| Number of New Contacts in Outgoing calls | This variable is defined as the number of calls made to new contacts; i.e. contacts that were not seen in the initial four weeks of the study but calls were initiated after the first four weeks. |

Privacy of user data was of utmost priority throughout this project. All data were secured and protected at standards applied to medical metadata. All metadata captured by the study clients was required to be no more detailed than those employed by typically installed apps like the Gmail and Instagram. The participants had the option to opt out of the studies at any time. The eventual goal is to design privacy apps that 'learn' user preferences over a short period of time (e.g. weeks) and can recommend privacy settings even after they stop receiving the data. Also, though outside the scope of the current work, the eventual privacy app coming out of this research project will be extended to run under the OpenPDS framework [17], [21]. OpenPDS (an open source personal data source) keeps personal data in the cloud under the purview of an individual user rather than the third parties like Google or Facebook.

Given that this is the first study to connect mobile phone metadata with privacy attitudes, we have adopted a multi-stage approach. We first tested multiple hypotheses based on existing literature connecting social behavior and privacy attitudes. We found multiple hypotheses to hold and significant associations between phone signals and privacy attitudes. This motivated the second stage of analysis where multiple features were combined into unified prediction models. The validation at each stage, yielded better confidence and interpretability for the next phase.

## 4       Hypotheses Testing

Existing studies [2], [13], [26] show maintaining privacy is a strong factor in determining how users present themselves and hence exerts an influence on their social interactions. For our study, we used the number of calls made by individuals over the period of study to determine their level of interaction. Past research has analyzed privacy and interaction on SNSs and has explored the relationship between privacy concerns and actual behavior on SNSs. For example, Acquisti and Gross (2006) found that one's privacy concerns were a weak predictor of the use of social network sites [1]. Based on these studies, we attempt to understand the relationships between greater phone interactions and privacy attitudes. We expect our study to show a negative relationship between number of calls and privacy attitudes.

**H1: Higher call count is associated with a lower concern for privacy**

**H2: Longer time spent on calls is associated with a lower concern for privacy**

Previous research [4], [8], [10] has shown the importance of maintaining social ties in today's world. Online SNSs support both the maintenance of existing social ties and the formation of new connections. While online social network sites offer an attractive means for interaction and communication, they also raise privacy and security concerns. Researchers [2], [12], [29] agree that maintaining these social networks require disclosure of personal information and encourage the sharing of information. Along with online social networks, phone calls are also a popular way to stay in touch. As the number of mobile phone users' increase, more and more information will be shared over phone calls. Individuals receive a number of phone calls outside of their network or known "friends and families" in a given day. These calls could be from marketing agencies, credit card sellers or even phishing, where criminals try to gain sensitive personal information using fraudulent means. In such a scenario, we believe that people who are highly concerned about their privacy would only respond to calls from within their network. Therefore, we hypothesize that a higher call response rate indicates a lower concern for privacy.

**H3: Higher call response rate is associated with a lower concern for privacy.**

**H4: Higher missed call rate is associated with a higher concern for privacy.**

The creation and maintenance of relationships is one of the chief motivations for an individuals' use of SNSs [1], [6]. Studies of the first popular social networking site, Friendster, [5], [11] describe how members create their profile with the intention of communicating news about themselves to others. The structure of these online social networks allows for the same information to be shared between close friends, strangers or acquaintances [1]. As newer contacts are included within an existing social network, there is more personal information shared within the network. Therefore, we hypothesize that people who have high privacy concerns will not readily initiate newer contacts. That is, a higher number of new contacts in a network indicates a lower concern for privacy.

**H5: Higher number of new contacts initiated is associated with a lower concern for privacy.**

Based on the data captured in this study, we tested the above-mentioned hypotheses, operationalized each based on the features defined in Table 1 and undertook Pearson's correlation analysis. As shown in Table 2, we found hypothesis H3, H4 and H5 to be statistically significant in the expected direction and H1, and H2 to be non-significant.

**Table 2.** Results of hypotheses testing based on data in Rutgers Wellbeing study for n=53

|    | Hypothesis Testing | Expected Direction | p-value | Pearson's Correlation Coefficient |
|----|--------------------|--------------------|---------|-----------------------------------|
| H1 | Not Significant    | --                 | 0.532   | -0.088                            |
| H2 | Not Significant    | --                 | 0.490   | -0.097                            |
| H3 | Significant        | Yes                | 0.043   | -0.279*                           |
| H4 | Significant        | Yes                | 0.017   | 0.425**                           |
| H5 | Significant        | Yes                | 0.025   | -0.313*                           |

Demographic analysis of the data also provided some interesting insights. Descriptive analyses of data from this study shows that while both males and females are concerned with data sharing, females tended to have a slightly higher concern for privacy with 60% females receiving a "very concerned" privacy score on Westin's Index while only 54% males received the same score. We had a diverse sample in terms of race and income groups with representations from Asian Americans, African Americans, Latino, and White populations. There were no significant variations in privacy attitudes across race and income groups.

Based on existing studies [1], [7], [13] analyzing information sharing and disclosure, we hypothesized that a higher call count and longer time spent on calls would be indicative of a low concern for privacy. However, our results showed that this was not the case. We found a significant correlation between the frequency of answering and not accepting calls and privacy attitudes. This implies that the calls individuals choose to answer on their smartphones may be determined by their attitude towards privacy and information sharing. Similarly, calls individuals choose to ignore or "miss" can be a reflection of their privacy concerns. Previous studies have analyzed information sharing and disclosure in online social networks, [12], [16] however, our results show that information sharing over smartphones maybe a significant predictor of privacy attitudes. There is a significant relationship between the number of new contacts included in an individual's network and their privacy attitudes. This implies that as an individual's network grows the amount of information disclosed increases. While similar results were found for online social networks, our study shows that networks formed over phone calls, can also be an important variable in determining an individual's information sharing patterns.

These preliminary results indicate that the interconnections between privacy needs and phone metadata are not yet fully understood but could yield interesting findings when analyzed systematically. This also implies that concern over privacy may not be apparent by an examination of the simplest features, but an analysis of more nuanced features, like how individuals react to phone calls in terms of the number of calls they actually respond to, may reveal a more interesting pattern.

A little over half (57%) the surveyed participants exhibited moderate (27%) to high (31%) concerns for privacy, and 43% participants had a low score or were "unconcerned" with sharing their data. This is very different from the results of the original survey conducted in 2003. Westin reported only 10% of the population was classified as "Unconcerned," with the majority of individuals displaying moderate (64%) to high concerns (26%) regarding the (ab)use of their personal information [38]. While our sample includes mostly single undergraduate students and is not representative of a larger population, such a vast difference in results, begs a question about the differences in information sharing attitudes of individuals in the early or mid 2000's to the present day.

## 5 Towards a Predictive Model of Personal Privacy Attitudes

Multiple significant hypotheses suggest predictive potential of nuanced phone usage metadata towards privacy attitudes. Hence we used the three features found to be significant in the analysis above to build a combined predictive model for privacy attitudes. We consider two different classifications for the privacy attitudes. First, is the conventional three category classification as suggested by Westin and second is a two-class categorization based on the median value split.

In the first scenario the classes were defined based on the criteria recommended by Westin as already described in Section 2. This resulted in a split as follows: Privacy Fundamentalists: 5, Privacy Pragmatists: 31, Privacy Unconcerned: 17, Total: 53. Given the multiple (>2) classes present we decided to use the MultiClass Classifier as implemented in Weka 3.6, with J48 decision tree as its underlying method. Further considering the relatively modest sample size, we decided to use Leave-One-Out cross-validation to tradeoff between the learning ability and the generalizability of the results. We also compare the proposed phone-features based approach with two other approaches. One is a baseline 'Zero-R' approach, which simply classifies all data into the largest category. The second approach is based on using Big-Five [20] personality variables, which have been shown by multiple efforts to be related with privacy attitudes [19], [26]. The same classification method was applied to the different approaches. Lastly, given the unequal size of the classes, we also report the ROC (Receiver Operating Characteristic – Area Under the Curve) statistic along with the accuracy scores. Multiple prior efforts have suggested ROC as a more interpretable metric for classification when dealing with unequal classes [9].

As shown in Table 3, the Phone-features based model performed better than both the compared approaches. Focusing on the ROC metric, the model yielded 36% better prediction than the baseline model. Contrary to the expectations, the results also suggest that personality based metrics may not capture the right kind of signals to have predictive ability on privacy attitudes.

**Table 3**. Classification results using different approaches for (a) three-way classification as per Westin's taxonomy and (b) two-way classification (High vs. Low Privacy Concern).

|  | Three-Way Classification | | Two-Way Classification | |
|---|---|---|---|---|
|  | Accuracy | ROC | Accuracy | Accuracy |
| Baseline (Zero-R) | 0.58 | 0.50 | 0.56 | 0.50 |
| Personality Features | 0.53 | 0.40 | 0.43 | 0.39 |
| **Phone-Usage Features** | **0.66** | **0.68** | **0.74** | **0.69** |

To ameliorate some of the complexities associated with multi-class (>2) classification, we also consider a two-way classification problem, where the classes were based on a median split. Multiple participants fell at the median score (8 out of 12) and this resulted in two roughly equal classes of sizes of 30 (below or equal to median) and 23 respectively. We ran the classification in Weka 3.6 using J48 decision tree algorithm with Leave-One-Out cross validation. As shown in Table 3, this resulted in accuracy of 74% at a two class classification task and an ROC metric of 0.69, which indicates a 38% improvement over the baseline. Again, the relatively poor performance of personality based features suggests that traditional personality type measures may not be suited to predict privacy attitudes. Further investigation with larger samples is needed to confirm this initial evidence.

## 6   Discussion

Limitations of this study include that our sample is from only one university and not from a nationally representative sample. The sample population was also not very diverse in terms of age as they were mostly undergraduate students between 21-23 years. Also, we used a self-report survey on privacy rather than observing and recording the participants' behavioral patterns in terms with respect to data sharing.

This study was carried out as an exploratory field study to understand how cell phone metadata can be used to build an individuals' personal privacy signature. While, we respect completely individuals' rights to their data, we posit that the current privacy debate is heavily biased towards the sharing and protection of socio-mobile data from third parties. Comparatively, little attention has been paid towards re-pivoting the same data to suggest privacy settings to users themselves for different applications. While similar studies have been conducted using online social network data, this study is the first to motivate and ground the use of phone metadata towards identifying the privacy attitudes and needs of individuals.

While the current work has focused on relatively simple set of features and tested a small number of hypotheses, the significant jump obtained in prediction ability points to the value in exploring this direction further. In particular the direction of using more nuanced behavioral features, over a correspondingly larger sample size and degrees of freedom is part of our future work. With appropriate refinements and advancements, the proposed methodology could allow for automatic privacy attitude understanding for billions of mobile phone users.

**Acknowledgements.** We would like to thank Cecilia Gal, Padampriya Subramnian, Ariana Blake, Suril Dalal, Sneha Dasari, and Christin Jose, for help with conducting the study and processing the data.

**REFERENCES**

1. Acquisti, A., & Gross, R. (2006). Imagined Communities: Awareness, Information Sharing, and Privacy on the Facebook. *Privacy Enhancing Technologies Lecture Notes in Computer Science,* 36-58.
2. Acquisti, A., & Grossklags, J. (0). Privacy Attitudes and Privacy Behavior. doi:10.1007/1-4020-8090-5_13
3. Beale, R. (n.d). Supporting social interaction with smart phones. Ieee Pervasive Computing, 4(2), 35-41.
4. Bourdieu, P., & Wacquant, L. (1992). Classification Struggles and the Dialectic of Social and Mental Structures. In *An invitation to reflexive sociology* (p. 14). Chicago: University of Chicago Press.
5. Boyd, D. (2006). Friends, friendsters, and myspace top 8: Writing community into being on social network sites.
6. Boyd, D., Ellison, N. (2007). Social Network Sites: Definition, History, and Scholarship. J. Comp.-Mediated Commun., vol. 13, no. 1, Oct. 2007, 210–30

7. Buchanan, T., Paine, C., & Joinson, A. N. (2007). Internet Privacy Scales. *Journal of the American Society for Information Science and Technology*.
8. Burke, M., Kraut, R., Marlow, C. (2011). Social capital on Facebook: Differentiating uses and users. *ACM CHI 2011: Conference on Human Factors in Computing Systems*
9. Chawla NV. Data mining for imbalanced datasets: An overview. *In Data mining and knowledge discovery handbook*, pp 853–867. Springer, (2005).
10. Coleman, J. (1988). Social Capital in the Creation of Human Capital. *Knowledge and Social Capital, 94*, 17-41.
11. Donath, J., & Boyd, D. (2004). Public displays of connection. *bt technology Journal*, *22*(4), 71-82.
12. Dwyer C, Hiltz SR, and Passerini K (2007) Trust and privacy concern within social networking sites: a comparison of facebook and MySpace. In: *Proceedings of the Americas conference on information systems 2007*, AIS, Keystone
13. Ellison N, Steinfield C, Lampe C (2007) The benefits of Facebook "friends": exploring the relationship between college students' use of online social networks and social capital. *Journal of Computer Mediated Communication* 12:1143–1168
14. Egelman, S., Oates, A., & Krishnamurthi, S. (2011, May). Oops, I did it again: mitigating repeated access control errors on facebook. In *Proceedings of the SIGCHI conference on Human Factors in Computing Systems* (pp. 2295-2304). ACM.
15. Fang, L., & LeFevre, K. (2010, April). Privacy wizards for social networking sites. In *Proceedings of the 19th international conference on World wide web* (pp. 351-360).
16. Fogel, J., & Nehmad, E. (2009). Internet social network communities: Risk taking, trust, and privacy concerns. Computers in Human Behavior, 25. doi:10.1016/j.chb.2008.08.006
17. Hang, A., Von Zezschwitz, E., De Luca, A., & Hussmann, H. (2012). Too much information! User attitudes towards smartphone sharing.
18. Introna, L., & Pouloudi, A. (1999). Privacy in the information age: Stakeholders, interests and values. *Journal of Business Ethics*, *22*(1), 27-38.
19. Junglas IA, Johnson NA, Spitzmüller C. Personality traits and concern for privacy: an empirical study in the context of location-based services. *European Journal of Information Systems*. 2008 Aug 1;17(4):387-402.
20. John, O. P., Naumann, L. P., & Soto, C. J. (2008). Paradigm shift to the integrative big five trait taxonomy. *Handbook of personality: Theory and research*, *3*, 114-158.
21. Karlson, A., Brush, A., & Schcchter, S. (2009). Can i borrow your phone? understanding concerns when sharing mobile phones. *CHI,* 1647-1650. doi:10.1145/1518701.1518953

22. Nordichi 2012: Making Sense Through Design - *Proceedings Of The 7Th Nordic Conference On Human-Computer Interaction*, 284-287. doi:10.1145/2399016.2399061
23. Reeder, R. W., Karat, C. M., Karat, J., & Brodie, C. (2007). Usability challenges in security and privacy policy-authoring interfaces. In *Human-Computer Interaction–INTERACT 2007* (pp. 141-155). Springer Berlin Heidelberg.
24. Schlegel, R., Kapadia, A.& Lee, A.. (2011). Eyeing your exposure: Quantifying and controlling information sharing for improved privacy. SOUPS 2011 - *Proceedings Of The 7Th Symposium On Usable Privacy And Security*, doi:10.1145/2078827.2078846
25. Squicciarini, A. C., Sundareswaran, S., Lin, D., & Wede, J. (2011, June). A3p: adaptive policy prediction for shared images over popular content sharing sites. In *Proceedings of the 22nd ACM conference on Hypertext and hypermedia* (pp. 261-270). ACM.
26. Steinfield, C., Ellison, N. B., & Lampe, C. (2008). Social capital, self-esteem, and use of online social network sites: A longitudinal analysis. *Journal of Applied Developmental Psychology*, *29*(2008), pp 434-445
27. Stutzman, F. (2006). An Evaluation of Identity-Sharing Behavior in Social Network Communities. *International Digital and Media Arts Journal*, 3(1), 10-18.
28. Rosenbaum, B., Attitude toward invasion of privacy in the personnel selection process and job applicant demographic and personality correlates, L. *Journal of Applied Psychology*, Vol 58(3), Dec 1973, 333-338.
29. Tufekci Z (2008) Can you see me now? Audience and disclosure regulation in online social network sites. *Bulletin of Science & Technology Studies* 11:544–564
30. Wang, Z., & Liu, Y. (2014). Identifying Key Factors Affecting Information Disclosure Intention in Online Shopping. *International Journal of Smart Home*, *8*(4).
31. Watson, J., Besmer, A., & Lipford, H. R. (2012, July). + Your circles: sharing behavior on Google+. In *Proc. of the Eighth Symposium on Usable Privacy and Security* (p. 12). ACM.
32. Westin, A. F. (1968). Privacy and Freedom. Washington and Lee Law Review.
33. Westin, A., and Harris Louis & Associates. Harris-Equifax Consumer Privacy Survey. 1991.
34. Westin, A., and Harris Louis & Associates. Equifax-Harris Consumer Privacy Survey. 1996.
35. Westin, A., and Harris Louis & Associates. E-Commerce & Privacy: What Net Users Want. 1998.
36. Westin, A. Freebies and Privacy: What Net Users Think. 1999 http://www.pandab.org/sr990714.html.

37. Westin, A., and Harris Interactive. IBM-Harris Multi-National Consumer Privacy Survey. 1999.
38. Westin, A. Consumer, Privacy and Survey Research. 2003 http://www.harrisinteractive.com/advantages/pubs/DNC_AlanWestinConsumersPrivacyandSurveyResearch.pdf